# TFPaaS : Test-first Performance as a Service to Cloud for Software Testing Environment[1]

Alim Ul Gias[2], Rayhanur Rahman, Asif Imran and Kazi Sakib
Institute of Information Technology, University of Dhaka
Ramna, Dhaka-1000, Bangladesh

**Abstract-** Performance Testing is critical for applications like web services and e-commerce platforms to ensure enhanced end user experience. In such cases, starting to test the system's performance early should significantly reduce the overall development cost. Test-first Performance (TFP) is one such paradigm that allows performance testing right from the early stage of development. Given such potential benefit, this paper proposes the design of a testing framework IVRIDIO which introduces TFP as a Service (TFPaaS). IVRIDIO incorporates the Plugin for TFP in the Cloud (PTFPC) aiming to provide instant feedbacks - a prime requirement of TFP to immediately fix critical performance issues. Furthermore, the Convention over Configuration (CoC) design paradigm has been applied by introducing a configurable project template to maintain TFP test cases. The prototyping details of the framework are given and the variation of response time to the inclusion of PTFPC has also been discussed. The Summated Usability Metric (SUM) score has been provided so that it can later be used for comparing the PTFPC plugin's usability.

## I. Introduction

Testing the performance of load-extensive applications (web services, e-commerce platforms) from the early stage can significantly reduce the overall development cost [1]. However, testing the performance right from the beginning will require an approach similar to Test Driven Development (TDD) [2]. Test-first Performance (TFP) [1] is one such paradigm which incorporates performance testing in TDD. TFP consists of three sub-processes- test design, critical test suite execution and master test suite execution. The tests are designed in advance by a performance architect who is not involved in coding but has close interactions with the developers and has a deep knowledge of the end users' expectations regarding the performance of the system. In order to get fast feedback, a critical test suite is designed to provide early warning of performance problems. However, to understand the overall system's behavior under heavy workloads a master test suite is designed and executed after the whole system is being implemented.

W. Jun et al. [3] have raised some important research questions regarding cloud testing that include - Why do we require testing in the cloud? As an answer they have stated that the overall development cost can be effectively minimized by migrating tasks like adaptability, stress and performance testing into the cloud. Since TFP is a nothing but a specialized approach for performance testing, organizations could get motivated for using the resources offered by the cloud for executing performance test cases. Cloud service providers should keep the fact in mind that critical test suites of TFP approach will be executed alongside the software development phase. Thus existing cloud testing services should provide certain components that will work along with the developers' IDE and be able to give instant feedbacks after executing the test cases.

Although multiple researches have been conducted on cloud based testing frameworks, to the best of the authors' knowledge those works do not explicitly address TFP. State of the art frameworks focus on generic architecture [4] [5] and specific issues like efficiency [6] , automation [7] [8], communication protocol [9], etc. For example, Cloud9 [6] is a testing framework that achieves significant amount of efficiency by using symbolic execution [10]. D-Cloud [7] [8] is another example which automates the testing procedure using descriptions of the system configuration and test scenarios. AGARIC [9] on the other hand, introduces the Test Flow Control Protocol (TFCP) for the interaction within different nodes of the service. However, it is arguable that existing frameworks comprise all the essential components required to support a particular framework like TFP.

We propose the design of a cloud testing environment IVRIDIO that extends D-Cloud [7] and AGARIC offering TFP as a Service (TFPaaS). To address the inadequacy of supporting TFPaaS, two sets of components have been modeled. The first set of components is at the clients' end which includes the Test Script Validator, TFP Service Communicator and TFP Application Identifier. These components are bundled together and named as the Plugin for TFP in the Cloud (PTFPC). The software design paradigm Convention over Configuration (CoC) [11] has been utilized in PTFPC to make the service availing procedure simpler by providing a configurable project template for maintaining TFP test cases.

---

[1] The initial version of this work is reported in INTECH 2013 [20]
[2] Corresponding author: alimulgias@gmail.com

The second set of components is at the TFP Service's end that includes the Test Receiver, Test Script Parser and Test Task Manager. Although these components persist in state of the art frameworks, their functionality has been adapted to support the TFP scheme. The service will receive test cases from the PTFPC and deliver results by means of SOAP [12] messages. This coordination within the PTFPC and TFP Service will ensure that the clients' are receiving instant feedbacks to readily address certain performance issues.

The framework was prototyped by dividing the system into three subparts to perform certain investigation. An experiment was conducted to see that whether the inclusion of the Test Script Validator in PTFPC has any effect in the overall response time of IVRIDIO. The results suggest that the variation is insignificant considering regular test tasks. Moreover, a Summated Usability Metric (SUM) [13] score of -0.4175 is obtained by the PTFPC which was calculated by assigning five software development professionals to complete a set of tasks using the plugin. This score can be used later to compare the usability of other client-side plugins that may be attained from future researches.

The rest of the paper is organized as follows: Different frameworks for testing in the cloud are discussed in Section II. Section III describes the proposed cloud testing framework IVRIDIO clarifying how it offers TFPaaS. Section IV presents the implementation details for prototyping the system and experimental results. A discussion on the experimental parameters is presented in Section V. Finally, this paper is concluded with future research directions.

## II. Related work

A prototype of Software Testing as a Service (STaaS) on cloud has been developed by Lian Yu et al. [4]. The researchers have evaluated the scalability of the platform by increasing the test task load. An analysis was presented regarding the distribution of computing time on test task scheduling and processing over the cloud. Moreover, they have compared the performance of their proposed algorithm to existing schemes. Their work have addressed four major issues- firstly clustering tenant task requests, secondly scheduling those clustered tasks, thirdly monitoring testing resources and task status, and fourthly managing cloud processes, processors and virtual machines by means of a dynamic approach.

Cloud-based Performance Testing System (CPTS) for web services has been presented in [5]. The CPTS architecture consists of three layers that include- user management layer, system control layer, and cloud computing service layer. The user management layer is responsible for providing the front end to handle user interactions. The functionalities of system control layer include- firstly managing and distributing performance test tasks, secondly making the test distribution and test run-control function available, thirdly managing test scripts, data, results, and lastly controlling the whole platform. The cloud computing service layer consists of the cluster server and the node server on which the execution of every process will take place.

Cloud9 is a testing service that allows to upload and test applications rapidly as a part of the software development life cycle [6]. It achieves noteworthy level of automation using symbolic execution [10]. This technique can explore all feasible execution paths in a program, and thus is an ideal candidate for test automation. Multiple challenges were addressed for improving the scalability of Cloud9's parallel symbolic execution engine. The challenges include- path explosion, constraint solving overhead, memory usage, the need to do blind-folded work partitioning, distributing the search strategy without coordination, and avoiding work and memory redundancy.

A testing environment named D-Cloud is presented in [7] [8] that uses the cloud computing technology and concentrates on an important fact that highly dependable systems must be fault tolerant for failure concerning both software and hardware. D-Cloud uses the virtualization aspect of the cloud to test hardware fault tolerance and it automates the testing procedure by using system configuration and test scenario descriptions to execute the tests. This methodology can also be used to improve the efficiency of other testing frameworks as it allows to automate the test case execution.

AGARIC [9] is a hybrid cloud based software testing platform which uses both centered cloud resources and scattered user owned resources to establish the test network. AGARIC comprises of three types of node which include- test control node, test center node, and test daemon node. Researchers involving the design of AGARIC have also proposed the Test Flow Control Protocol (TFCP) to organize diversely distributed test daemon nodes to a resource pool and to communicate among those three nodes. The protocol uses stateless HTTP as a base and a basic set of actions is used for conducting the communication.

The review of the existing frameworks illustrates that though several issues have been considered in designing a software testing environment in the cloud, none of the framework offers a service based on instant feedback which is essential for a testing paradigm like TFP. For having a seamless execution of the TFP test case, clients' end must not be overlooked and new tools should be provided to developers for effective integration of TFP alongside the regular development phase. IVRIDIO is one such framework which offers TFPaaS and ensures proper execution of TFP by introducing a set of new components that will work along with the clients' development environment.

## III. TFPaaS via IVRIDIO

IVRIDIO is a software testing framework that focuses on offering TFPaaS by means of cloud computing technology. The architectural components of IVRIDIO are bound to provide the service using defined interactions within themselves. In addition, the CoC is being applied during the design of IVRIDIO to simplify the service availing procedure. This section aims to clarify those details of IVRIDIO with an assumption that a web service, which demands high performance, is being tested. Besides, an overview of the TFPaaS is given to understand the sequence of actions needed to avail the service.

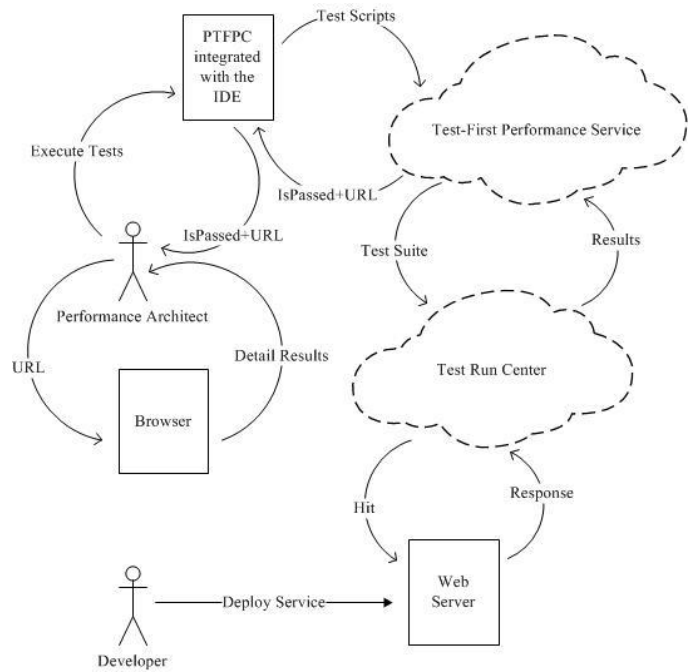

Figure 1: Top level View of IVRIDIO

### A. System Architecture

The top-level overview of the proposed architecture consisting of three core components is shown in Figure 1. The first component is the PTFPC which is designed to be assimilated with the IDE of the clients. PTFPC is designed by centering on the fact that it will aim to provide a seamless execution of the critical test suite. Such execution will be provided alongside the regular software development by providing instant feedbacks required to fix different performance problems. The instant feedback will also ensure the location transparency (users cannot tell where resources involving test script execution are located) which is essential in distributed systems like the cloud.

The second component is the Test-first Performance Service (TFPS) which converts test scripts to a runnable set of test suites. These test scripts are provided to the TFPS by the PTFPC when it receives the command for executing test cases from the performance architect. Performance architect does so, as soon as a developer deploys a specific service to the web server. The third component is the Test Run Center that executes runnable test suites provided by the TFPS. Results which are obtained after executing test suites are returned to the service. The service will store those results in a repository and a summarized version will be provided to performance architect via PTFPC. A URL will also be given for viewing the detailed results by using a simple web browser.

The component stack of the proposed framework is represented in Figure 2. Components having thick borders are core elements which aid in offering TFPaaS and will be focused in this paper. These components are either newly introduced or modified from existing frameworks like D-Cloud and AGARIC. Other components are supporting elements and similar to elements of [4] and [5].

The top layer of the stack is composed of User End components and comprises the newly introduced PTFPC. The plugin is subdivided into three parts. The first component is the test script validator that will run on the background and continuously validate test scripts which are being written. In existing frameworks this component resides in the cloud. However, as validation process does not demand huge resources offered by the cloud, it can be moved to the user end. Besides, as the validator will continuously run on the background, testers will be aware of their mistakes instantly when they are writing test scripts. Moreover, a tester will not like to get a feedback from the cloud that there is an error in the test script, rather it is better to identify the error right from her/his own machine.

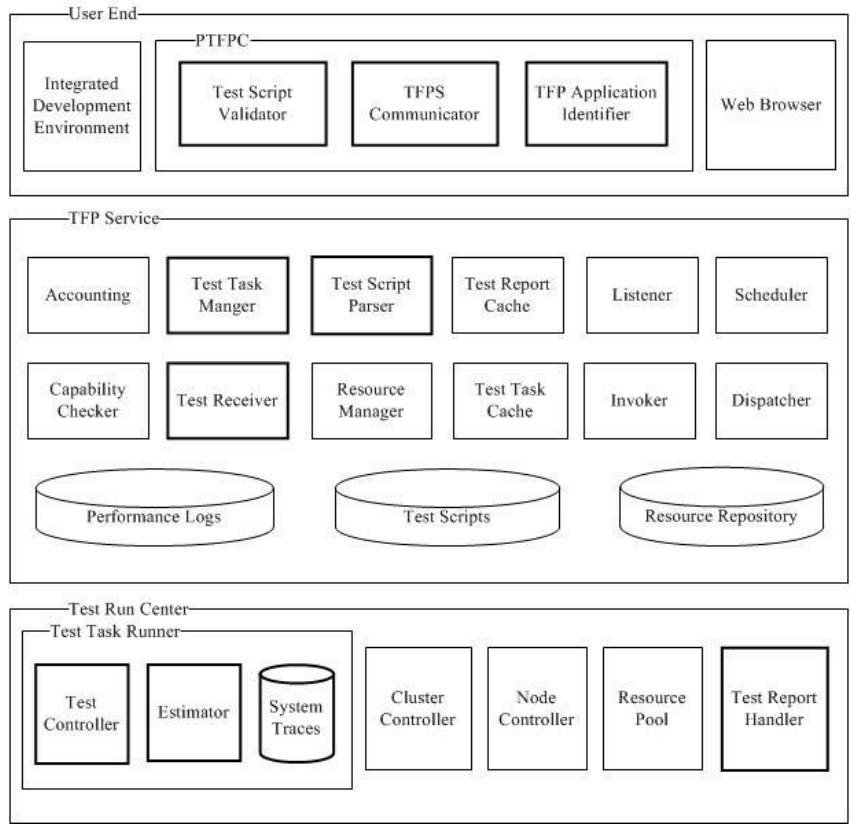

**Figure 2: The component stack of IVRIDIO**

The TFP Service (TFPS) Communicator is the second component of the plugin and will handle the communication with the TFP Service. To be more precise, the communicator will forward test scripts to the service and receive results after the test case execution. Test scripts can be written following the pattern of scripts in state of the art framework like D-Cloud. As shown in Figure 3, the validator will provide a validated test script to the communicator in time of creating the SOAP message.

The third component- TFP Application Identifier will provide an application identity which will be stored in the TFP Service's database to track the specific application. Besides, it will also provide user credentials and for that, the user has to be logged in into the IDE. After receiving the required data from the validator and identifier, the communicator will create a SOAP message and send it to the TFPS.

The generic structure of a SOAP message for communicating with TFPS is shown in Figure 4 having three subparts. The first part is for the identification of the application in the cloud's end. The second part consists of the test case designed in format similar to D-Cloud. The last part consists of the performance criteria based on which it will be decided that whether the service meets certain performance standards. Description of the message's child nodes that are related to the test case is provided in Table 1.

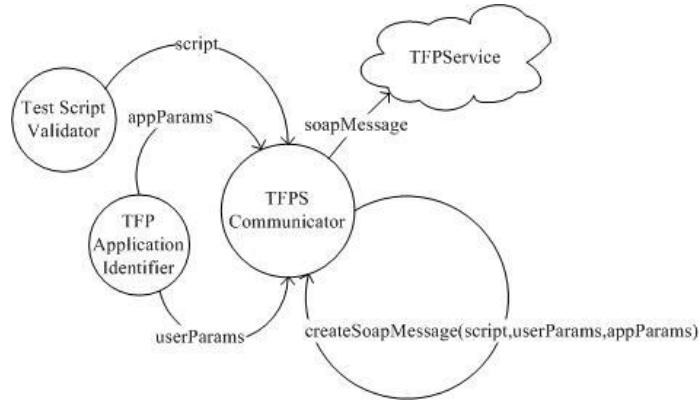

**Figure 3: The interaction within the three components of the PTFPC**

```
<?xml version="1.0"?>
<soap:Envelope xmlns:soap="http://www.w3.org/
2003/05/soap-envelope">
  <soap:Header>
  </soap:Header>
  <soap:Body>
    <m:TFPService xmlns:m="URI">
      <m:application>
         <m:appId></m:appId>
         <m:userName></m:userName>
      </m:application>
      <m:case>
         <m:url></m:url>
         <m:method></m:method>
         <m:message></m:message>
      </m:case>
      <m:criteria>
         <m:response></m:response>
         <m:tps></m:tps>
         <m:bps></m:bps>
      </m:criteria>
    </m:TFPService>
  </soap:Body>
</soap:Envelope>
```

**Figure 4: The generic structure of a SOAP message for communicating with the TFP Service**

**Table 1: Description of the test case attributes within the SOAP message**

| Attribute | Description |
|---|---|
| appId | The application identity which will be provided by the PTFPC Application Identifier |
| userName | The username of the current logged in user into the IDE |
| url | The URL of the web service which will be under test, for example www.example.com/TFP/ |
| Method | HTTP GET or HTTP POST |
| Message | The message that should be posted on the provided URL |
| Response | The expected response time, for example 3 millisecond |
| tps | The expected throughput i.e. transactions per second, for example 30 |
| Bps | The expected bits per second, for example 1048576 |

The middle layer of the proposed framework is the TFP Service layer. Three major components, focusing on an efficient execution of test cases are the Test Receiver, Test Task Manager and Test Script Parser. Figure 5 illustrates interactions among those components which will take place after receiving a test case by the Test Receiver. The receiver will provide the SOAP message to the test manager in a XML format. The test manager will forward this file to the test script parser which will parse the file to provide a set of instructions and application identity to the manager.

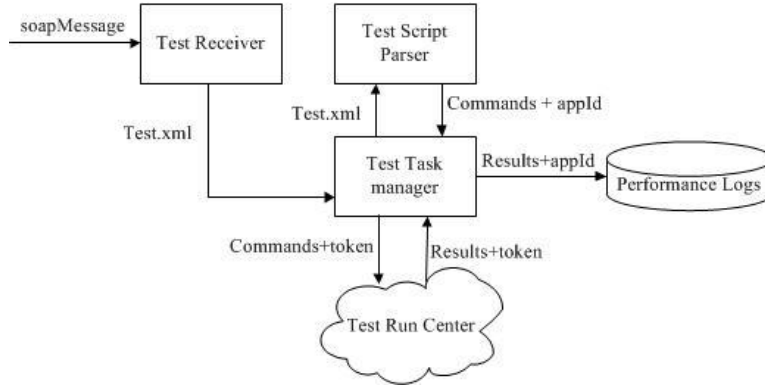

**Figure 5: The interaction among the three core components of TFP Service**

After receiving instructions, the manager will send those to the final layer - Test Run Center via protocol similar to TFCP [9]. The master test suite can be executed adaptively to provide better results which can guide to find bottlenecks of the application under test. However, for that the web services have to be deployed within IVRIDIO's execution environment enabled to keep track of system's traces. This adaptive software testing can be implemented using the guidelines provided in [14]. The components which are responsible for this task are the Test Task Runner and Test Report Handler. Test Task Runner has three components which are – Test Controller, Estimator and a log for keeping system traces.

An illustration of the adaptive testing procedure is shown in Figure 6. The test controller will pass the commands to the execution environment and traces of those executions will be stored. These traces will be used by the estimators to provide guidelines for performing the test in the next iteration. The results if the test script execution will be sent to the report handler who will forward it to the test task manager of the TFP Service. The result will be stored in the database and the notification will then be provided to the PTFPC.

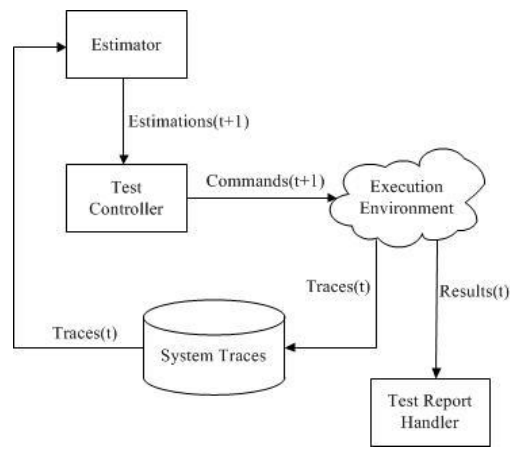

**Figure 6: Illustration of Adaptively Executing the Master Test Suite**

## B. Convention over Configuration

The principle of CoC design paradigm is to reduce the humanly decision to gain simplicity without losing the flexibility. This principle is incorporated during the design of IVRIDIO to reduce complexities of availing TFPaaS. This was accomplished by providing a project template along with PTFPC. Given that the template is followed, the plugin will be able to reduce number of steps required to execute the test case for a specific service. As IVRIDIO is a cloud-based testing framework it will certainly help to reduce the overall development cost [15]. Nevertheless, the benefit provided by CoC in the context of a large project will further reduce development cost due to the fact that it is removing certain human involvements from the testing procedure.

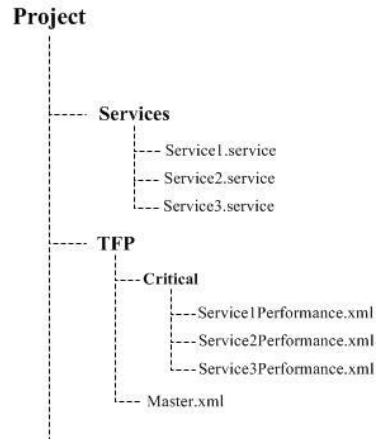

**Figure 7: The directory structure according to the project template provided**

The project template provides the directory structure shown in Figure 7. It can be observed that the name of a test case is given in the format *<ServiceName>Performance*. All critical test scripts are kept under the directory *TFP\Critical* and the master test script is kept in the *TFP* folder. This directory pattern eliminates the task of binding each service with its test case and thus the PTFPC will be able to identify and execute test cases for each service without any configuration file being written. However, the flexibility will be kept such that the developer will also be able to change this directory structure, according to their own convenience, by maintaining a configuration file.

The directory pattern will yield multiple other benefits if certain features could be ensured during the implementation. The pattern should allow users to create the test case file corresponding to the service by simply right-clicking on the service file and then giving the command to create critical test case file. Similarly, test cases can be executed by right-clicking in the service file and giving the command to run critical test case. However, these options will not be applicable for the master test suite as it does not require frequent execution and is not related to one specific service. Although, feature can be added such that the master test case file will be created whenever a TFP project is being opened. Moreover, options can be added to execute the master test suite by right-clicking on the project root-folder and then giving the command to run master test suite.

## C. Test-First Performance as a Service

IVRIDIO provides TFPaaS using the algorithm provided for TFP by Johnson et al. in [1]. Initially performance test cases should be designed and after the design completion the developer has to implement a functionality. The implementation will continue until it passes its functional test. Upon passing the functional test, the service will be deployed in a web server. These preliminaries are illustrated in Figure 8.

Certain steps that should be performed by the PTFPC to send a test case to the TFP Service are shown in Figure 9. The performance architect will give the command to execute test case for the corresponding service. PTFPC will sort out the respective test script and forward it to the TFP Service. If the test case for the service is not written then an error message will be generated.

The service will receive the test script as shown in Figure 10. After the script being received, it will be parsed into a set of instructions. These instructions will be forwarded to the test run center for execution. Upon executing those instructions, a performance log will be generated which will later be used for judging the performance of the web service.

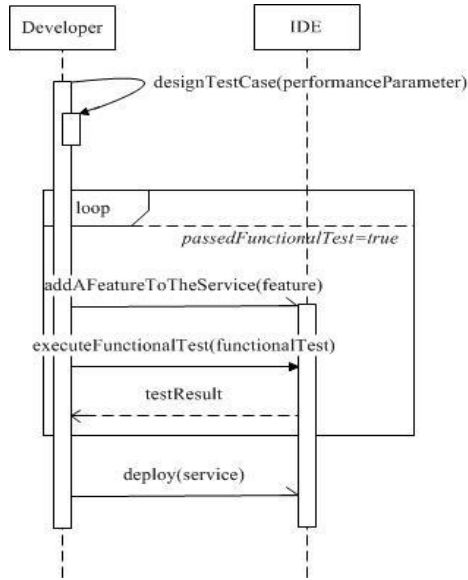

**Figure 8: A sequence of actions showing how a functionality will be implemented using the TFP principle**

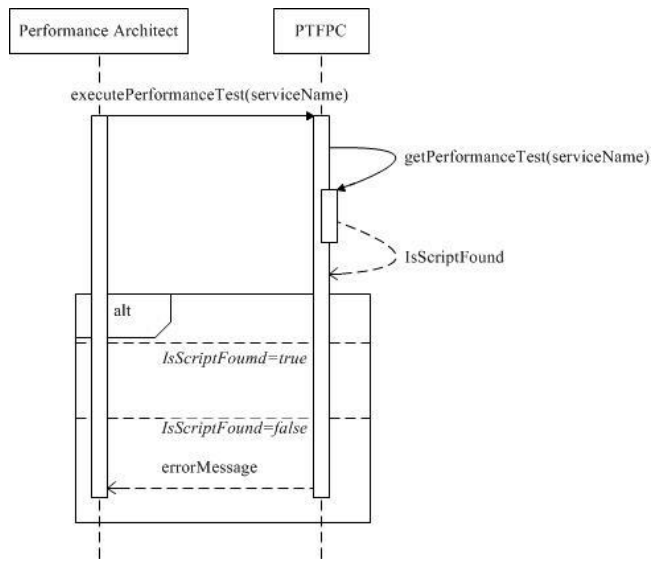

**Figure 9: Sequential steps followed by PTFPC to send a test case to TFP Service after getting the service name**

TFPaaS offered by IVRIDIO ensures that the location transparency is established. This can be understood by observing the above mentioned procedure to avail TFPaaS. The developers will pass the command to execute the test case from their IDE and the rest of the task will be handled by PTFPC. Moreover, instant feedbacks will be provided after the test case execution, ensuring that the developers will feel like working within the same environment. Achieving this is an essential requirement for any distributed system.

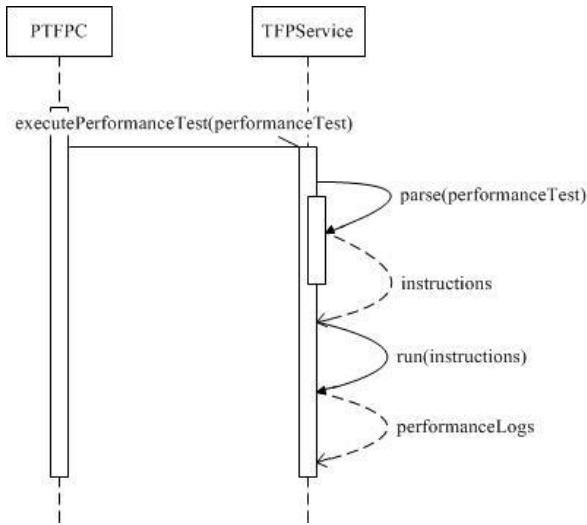

**Figure 10: Sequential interaction between the PTFPC and TFP Service for executing a test case and storing the result**

## IV. Experimental Setup and Results

This section initially provides a sketch for prototyping IVRIDIO due to certain experimental purposes. Since the test script validation part of IVRIDIO will reside on the client's end, an experiment was conducted to measure its effect on the overall execution time. Moreover, as the PTFPC will reside along the developer's IDE, it is important to get a measure of its usability. The Summated Usability Metric (SUM) score proposed by J. Sauro et al. [13] was used in this case to get an approximation of PTFPC's usability.

### A. Prototyping

The framework can be implemented easily if it is considered as three separate components interacting within themselves. These three components are PTFPC, TFPS and Test Run Center. The implementation of PTFPC may differ based on the IDE upon which it will reside. Although the implementation of the TFPS and Test Run Center may also differ based on the used technology, they should be deployed in a public and private cloud environment respectively. A possible scheme to implement these three components is illustrated in Figure 11.

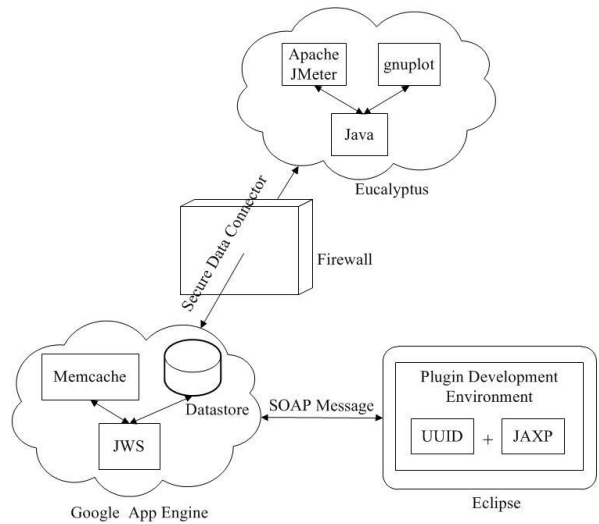

**Figure 11: Implementation scheme of IVRIDIO**

PTFPC was implemented for the Eclipse IDE using its Plugin Development Environment (PDE). The application identifier was developed with Universal Unique Identifier (UUID) and Java API for XML Processing (JAXP) was used for handling test scripts. TFPS was deployed in the public cloud environment - Google App Engine (GAE)

[16]. GAE provided the platform to build the service using Java Web Services (JWS) API. Besides for cache and storage purpose, GAE can provide the Memcache Java API and Datastore feature respectively.

The test run center was separately implemented inside a private network however it can also be developed in a private cloud using open source software Eucalyptus [17]. GAE can provide the Secure Data Connector (SDC) which allows data communication from a private cloud behind a secure firewall to a public cloud like TFPS. Apache JMeter was used as the performance testing software as it has the advantage of easily being installed in a distributed system like the cloud. For providing a graphical representation of results, gnuplot was used and the test task runner was implemented using Java programming language.

## B. Response Time Variation

Two web services were developed where one following the design similar to TFPS did not include the test script validation module. On the other hand, another web service included a component to validate submitted test scripts. The first case replicates the validation process of IVRIDIO where test scripts will be validated on the back ground by the PTFPC, making the validation time tends to zero. Thus to test its impact, the response time of both web services for varying number of requests were recorded.

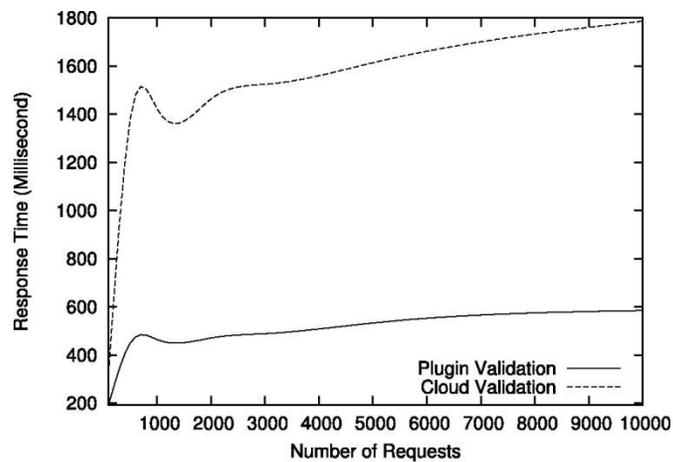

Figure 12: Graph showing difference in response time of validating test scripts with cloud and with plugin when the difference between total response time and validation time tends to zero

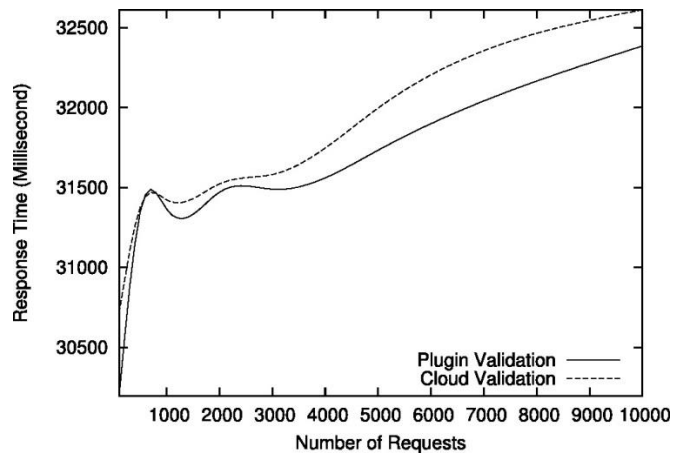

Figure 13: Graph showing difference in response time of validating test scripts with cloud and with plugin when the difference between total response time and validation time is much greater than zero

From the experimental data it is observed that if rest of the response time excluding the validation is very small, overall response time is much lesser which can be visualized from Figure 12. The graph is plotted by using the response time of each service for varying number of requests. The distance between the two curves is significant because the time required to complete rest of the tasks except validating scripts is kept much lower.

If we randomly increase rest of the time within a large range, the gap between the two curves will be almost negligible as shown in Figure 13. The later scenario is most likely to occur in real life that is validation time will be almost negligible considering rest of the tasks. Thus it can be said that moving the validation process from the cloud will not have much impact on the overall response time.

### C. SUM Score of PTFPC

The prototype of the PTFC was developed having two functionalities- creating and running a test case by clicking in the corresponding web service file. The summated usability of the prototype was calculated using the method described in [13]. This was particularly important as this score can later be used to approximate the comparative usability of PTFPC with other plugins getting developed in further researches. Five professionals including two Intern Software Engineer, two Senior Software Engineer and one Software Test Engineer were used to conduct the experiment. They were given a simple task of creating a certain number of test script files and then running those script files using the plugin.

The dimensions and metrics involving the SUM score calculation are provided in Table 2. Three ISO/ANSI dimensions were chosen and four metrics were selected to represent them. For calculating the task time we had initially set the ideal task time following the method proposed in [18]. The Z-Score was then calculated by subtracting the raw task time from the ideal one and dividing by the standard deviation. The error rate was calculated by counting the error opportunities and number of errors made during the testing. Considering the developed prototype of PTFPC we considered a single error opportunity – trying to run a test case without creating it. The number of error opportunities will certainly differ in the full implementation of the plugin. The Z-Score was calculated by dividing the total number of errors by the error opportunities.

Task completion was measured as the ratio of failed tasks to attempted tasks. This proportion of defects per opportunities has a corresponding z-equivalent that was looked up in the standard normal table. To calculate the Z-Score for composite satisfaction the average rating of a user's satisfaction score was subtracted from the mean rating of 1-5 scale [19] and divided by the standard deviation.

After calculating the four values, we assigned equal weights (0.25) and then sum those up to reach our SUM score of -0.4175. Although the scenario and implementation may differ, this value can be used as an approximation to evaluate the usability of similar plugins. For a meaningful comparison it will be better to not entirely relying on the concrete score. The comparative complexities of the system should be considered and certain factors (number of error opportunities, number of testers, number of trials, etc.) should be modified to reach into a conclusion from the score provided in this paper.

**Table 2: Summated Usability Metric Score of PTFPC**

| ISO/ANSI Dimensions | Metrics | Z-Score | Weighted Z-Value | SUM |
|---|---|---|---|---|
| Efficiency | Task Times | 1.58 | 0.395 | |
| Effectiveness | Error Rates | -3.49 | -0.8725 | -0.4175 |
| | Task Completion | 3.49 | 0.8725 | |
| Satisfaction | Avg. Satisfaction | -3.25 | -0.8125 | |

## V. Discussion

It is observed that moving the test script validation module from the cloud does not have a significant impact on the overall response time. However, it will certainly boost-up the overall usability of the system as the testers are able to fix script related errors instantly. However, performance script standards may vary in different organizations. Thus the service providers should define the allowed syntax for writing test scripts. The validator will work based on those syntax. The plugin must be synced with the test service so that it can readily receive further syntactical updates.

One of the steps of measuring task time was to find the ideal task time and subtracting each raw task time from it. If the task takes longer to execute the ideal task time will also relatively increase [18] making their difference almost similar in each cases. Thus the task time will have a minimal effect on the SUM score.

Testing the plugin with multiple set of testers may yield a more prominent result regarding average satisfaction. Although it can be assumed that if a set of testers are satisfied with the system, multiple set of testers' satisfaction level will not change to a great extent. Moreover, the satisfaction level may not vary for plugins providing more or less similar functionalities and so its effect on usability for different plugins should remain the same.

However, the usability will heavily rely on plugin complexities (steps required to complete a task) and total functionalities as these can easily increase the error opportunities. Moreover, it may significantly increase the ratio of failed tasks to attempted tasks and eventually affecting the overall usability score. Thus during the comparison of PTFPC's SUM score with a plugin of same nature, the complexity and functionality ratio should be analyzed for getting a better estimation.

# VI. Conclusion and Future Work

This paper introduces a testing framework named IVRIDIO that offers TFPaaS. IVRIDIO emphasizes on the client's end by adding the PTFPC that allows the developers to execute critical test suites along with the regular development phase. The PTFPC focuses on providing a fast feedback which is one of the prime requirements of TFP. The solution has the additional advantage of lower test costs since it is based on cloud platform. Moreover, the framework also incorporates the paradigm Convention over Configuration to illustrate its effectiveness in reducing overall test effort.

The prototyping details of IVRIDIO are provided and it can be used to implement the system for further research. Moreover, the variation of response time due to the inclusion of the test script validator in the clients end is also discussed. It is observed that the inclusion will have minimal impact on the overall response time considering generic performance test tasks. The SUM score of PTFPC is also provided so that it can later be used to approximate the usability of other similar plugins.

In IVRIDIO the test script validation part was moved to the user end to increase system's usability and investigate its effect on the overall response time. This raised a vital research issue that which components from the cloud's end can be moved to the client's end. It should be kept in mind that some processes require the huge computational resources of the cloud. However, for processes which do not require such luxury may be moved to the client end. Identifying those processes is a significant task, and still remains as an open research challenge.

Another future challenge which evolved from this research is designing a plugin that will support multiple cloud based testing services. The PTFPC proposed in IVRIDIO is explicitly designed to support TFP. However, certain issues will arise if someone wants to avail a service that offers another testing methodology. Thus a further research challenge is to design a standard architecture for a plugin of this kind.